# Data Life Cycle Labs

## A New Concept to Support Data-Intensive Science

Jos van Wezel, Achim Streit, Christopher Jung,
Rainer Stotzka, Silke Halstenberg, Fabian Rigoll,
Ariel Garcia, Andreas Heiss
Karlsruhe Institute of Technology (KIT)
Karlsruhe, Germany
Achim.Streit@kit.edu

Martin Gasthuber
Deutsches Elektronen-Synchrotron (DESY)
Hamburg, Germany
Martin.Gasthuber@desy.de

Kilian Schwarz
GSI Helmholtzzentrum für Schwerionenforschung GmbH
Darmstadt, Germany
k.schwarz@gsi.de

André Giesler
Forschungszentrum Jülich (FZJ)
Jülich, Germany
a.giesler@fz-juelich.de

*Abstract*—In many sciences the increasing amounts of data are reaching the limit of established data handling and processing. With four large research centers of the German Helmholtz association the Large Scale Data Management and Analysis (LSDMA) project supports an initial set of scientific projects, initiatives and instruments to organize and efficiently analyze the increasing amount of data produced in modern science. LSDMA bridges the gap between data production and data analysis using a novel approach by combining specific community support and generic, cross community development. In the Data Life Cycle Labs (DLCL) experts from the data domain work closely with scientific groups of selected research domains in joint R&D where community-specific data life cycles are iteratively optimized, data and meta-data formats are defined and standardized, simple access and use is established as well as data and scientific insights are preserved in long-term and open accessible archives.

## I. Introduction

Today data is knowledge – data exploration has become the 4th pillar in modern science besides experiment, theory, and simulation as postulated by Jim Gray in 2007 [1]. Rapidly increasing data rates in experiments, measurements and simulation are limiting the speed of scientific production in various research communities and the gap between the generated data and data entering the data life cycle (cf. Fig1) is widening. By providing high performance data management components, analysis tools, computing resources, storage and services it is possible to address this challenge but the realization of a data intensive infrastructure at institutes and universities is usually time consuming and always expensive. The introduced "Large Scale Data Management and Analysis" (LSDMA) project extends the services for research of the Helmholtz Association of research centers in Germany with community specific Data Life Cycle Laboratories (DLCL). The DLCLs are complemented with a Data Services Integration Team (DSIT) which provides generic technologies and services for multi-community use based on research and development in the areas of data management, data access and security, storage technologies and data preservation.

The LSDMA project initiated at the Karlsruhe Institute of Technology (KIT), builds on the familiarity with supporting local scientists at a computer center, the knowledge of running the Grid Computing Centre Karlsruhe (GridKa) [2] as the German Tier 1 hub in the World Wide LHC Computing infrastructure [3], the Large Scale Data Facility (LSDF) [4] and the experience with the very successful Simulation Labs [5] that specialize at supporting HPC users.

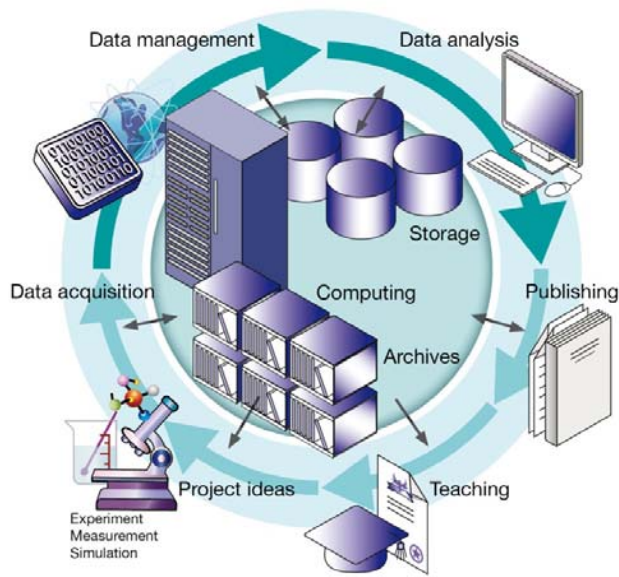

Figure 1.  The scientific data life cycle

The project partners from the Helmholtz Association have an equal long standing experience in large data and computing support in various communities. LSDMA brings together the combined expertise from all partners as well as the data intensive research communities that are local at their institutes.

## II. LARGE SCALE DATA IN MODERN SCIENCE

Modern science and scientific computing is about data. In the process from collecting data till publication, the data has been moved, aggregated, selected, visualized and analyzed. In view of the ever increasing amounts of data this process must be organized and structured. Data management is the organization and structuration of the data life cycle which will allow faster results and dependable long term references. The data life cycle does not stop after publication. Without a suitable data organization or infrastructure (e.g. with persistent identifiers) data can no longer be found and is essentially lost. Data has become a source of knowledge in itself [6]. By combining, overlaying, mining, visualizing and application of various analysis methods the scientific process is now also driven by the content of already existing data collected in experiments and measurements. Data from every section of the data life cycle, including the raw measurement or simulations and data from rare and unique events, has become a valuable good that needs to be preserved and is lost if not properly managed.

Common in many experimental sciences are the vast requirements for large scale storage, data management and analysis resources. The actual requirements of the communities must be examined in detail because we have found a large variation of data management awareness and experience across communities in our discussions with them. Some scientific disciplines already have local solutions and are investigating e.g. the possibility of data sharing within the community whereas others are struggling with basic data handling and have only recently begun to organize their unstructured data.

### A. Data and analysis in climate and environment

#### 1) Atmospheric research

Our understanding of the atmosphere is fundamentally limited by the sparseness of highly resolved regional and global observations. Data from remote sensing instruments like GLORIA [7] and MIPAS [8] on the ENVISAT satellite [9] promises to improve this situation substantially. Both instruments require data storage capacities of several hundred Terabytes per year (raw and processed data). Further projects and instruments are on the horizon, requiring similar amounts of data storage and data management and analysis techniques. The raw data consist of interferograms with a very high spectral resolution that are converted in multiple steps resulting in many (> 10000) small files per day. During processing the access of such a great amount of small files is the limiting factor compared to the computing times for analyzing the data. Due to the irreplaceability and preciousness of the data, they must be archived for several decades

#### 2) Climate modeling

Climate model data from HPC simulations is being used worldwide by a growing number of users. The amount of new data ranges from 10 to 100 PB per year in international modeling activities (e.g. CMIP5 [10] and CORDEX [11]) which are archived and disseminated in globally distributed data federations. In recent years data applications are changing from community specific to interdisciplinary applications. Data federations have been established and the scientific research is treating more and more coupled and interdisciplinary questions like adaptation and mitigation. After annotation and quality control, around 25% of the climate model raw data move into long-term data archives. These model data will not be reduced anymore because usually detailed use cases are unknown in advance and therefore it is not predictable which outcome will be necessary for further investigations. After passing quality control the data is freely available and subsequent access is very unpredictable. Worldwide available archives for climate data must have ample capacity and must assure long term availability as well as related data curation services [12].

Data discovery, efficient data access, optimizing the processing workflows, security and quality management are only several examples which have to be considered for optimizing the data life cycle of climate model data. A fast und secure data access and transfer is also an essential requirement of the community for an interdisciplinary use of the archived data in international federations.

### B. Data and analysis in energy research

Data intensive computing in energy research comprises many topics that are driven by public interest and reflects the prominent international role Germany plays in the field of renewable energy. Applications at KIT focus on the area of energy generation, intelligent distribution, energy storage and intelligent management. Very large computing requirements in energy research originate from fusion research . In preparation of the international ITER [14] cascades of multi-scalar simulations are run to improve the fusion physics model and the condition predictions of the running reactor. The ITER project also uses distributed computing which implies the need for high speed and reliable data exchange between supercomputers. The experiments such as the large scale parameter studies of plasma analysis must be able to store and track the intermediate and final result of the computations as well as the associated meta-data in which the simulation conditions are kept for future reference. Data must be available for decades in the future since the ITER project is expecting its first demonstrations not earlier than 2040.

In the domain of energy distribution across a country, profiling and analysis of the consumption of electric energy using customer data to optimize the construction and use of the electric power grid is state of the art and often referred to as smart grids [15]. Originally to optimize the efficiency and reduce wear of power generation plants smart grids are increasingly expanded with smart metering in households. The steepness of the power curves is influenced by intelligently monitoring, predicting and controlling the consumption in the power grid and of the power generation in classic plants and increasingly, from alternative sources i.e. solar and wind. To be effective and efficient the analysis results of large scale data acquisition and simulation must be available quickly and in high detail. The amount of data to be stored and analyzed from power metering at main and sub circuits quickly reaches several TBs even for low data rates in small grids [16].

In battery research [17], theory and experiment are combined to improve the understanding of the internal atomic and molecular processes leading to more efficient and economic storing of energy which is an important prerequisite for future electric mobility. Application specific model systems are investigated using theoretical computations and measurements. Data analysis from specific in-situ measurements and multi-scalar modeling result in new detailed insights of the process flow under different reaction conditions. The experimental and theoretical quantum mechanics computations result in hundreds of GBs per computation that must be annotated and archived. This research field also has ample requirements for automatic data analysis workflows.

*C. Data and analysis in medicine*

In the human brain over 50 billion nerve cells communicate with each other, forming a network of unimaginable complexity [18]. The functionality of each nerve cell depends on its molecular structure and mechanism but also on its location within the brain [19]. Scientists of different disciplines cooperate to unravel these dependencies using different imaging techniques i.e. examining histological brain sections using modern scanning microscopes, structural and functional magnetic-resonance tomography (MRT), positron emission tomography (PET) supplemented with analysis methods from micro anatomy, cell biology and genetics as well as from physics and informatics. The ultimate goal is the compilation of a multidimensional virtual human brain atlas in which structures at the molecular level are linked to spatial functionality. By understanding the healthy brain the next step can be taken by distinguishing differences in people suffering from neurological or psychiatric disorders.

The construction of the virtual brain begins with 50nm thick tissue samples in which molecules, cells and structures are identified using state of the art image analysis techniques. This data is then used for a 3D and higher connectivity reconstruction. The process also recognises individual differences which are used to develop a probabilistic brain map. In this case the analysis of ten brains is used to correlate variability of size and location of different known regions. Next to the spatial functionality researchers also study the interaction of different regions thereby trying to resolve the relation of structure and function. For every brain about 1500 slices are analysed using different methods resulting in cytoarchitectonic and molecular maps on one hand, and to establish the relationship between function and connectivity on the other. For each brain slice around 18 polarization microscopy images are taken, resulting in 0.5 TB data. All images together amount to almost 1 PB of data which forms the basis for an anatomical correct 3D representation.

The data acquisition from a single brain takes several years. Subsequently the data has to be kept for an indefinite time and must be made internationally available for sharing by hundreds of researchers. Additionally this research field is to be supported with the handling of large scale data sets, the linking of the image data in complex post-processing computations as well as the development of meta-data catalogues and distributed storage and archives.

*D. Imaging in biology and materials research*

One of the compelling reasons for the fast increase of data management requirements is the use of digital imaging techniques for screening large populations of a species. Data handling and image analysis vary for each community, are different for each imaging technology and are continuously under development. Standard data policies for the data life cycle as well as common tools do not exist. The data life cycle lab for key technologies has started support and development of re-usable, efficient and automated data management and analysis tools for the ANKA synchrotron [20] and for the image processing of experiments using nanoscopy, electron microscopy and single plane illumination microscopy.

Spectroscopic and Röntgen diffraction techniques at the ANKA synchrotron are used for materials research and the clarification of biological structures. Unique for Germany, ANKA operates a beam line for steady-state and time-resolved synchrotron radiation circular dichroism, a technique which can be used to investigate structures of proteins and other macromolecules in a wide range of experiments e.g. for the determination of the secondary structure of proteins that cannot be crystallized, the investigation of the effects of the binding of ligands on protein secondary structure and cross membrane transports.

Different beam lines will shortly produce hundreds of TB per year. A huge variety of experiments e.g. biological high time resolution in vivo experiments, phase flows and combustion processes are recorded with up to 1200 samples/s, followed by time resolved volume reconstruction of the objects. The development of a common data management system applicable for all experimental setups has not been given a high priority this far and their current workflows are missing data and meta-data organization schemes, federated user management, long-term archives and common data transport and access [21].

In nanoscopy [22] photo activated localization microscopy is used to reach a tenfold resolution beyond ordinary light microscopy limits. In principle this allows imaging of subcellular structures but the technique is still under development and on-line efficient analysis is not available. Nevertheless, many medical and biological questions center on nano-scale structures e.g. those of DNA and DNA repair activities. Development of a workflow that integrates data management and fast analysis for many cells simultaneously and with high resolution will allow better insight into live cellular processes.

The digital representation of a single cell (20 µm) in nanoscopy, at a resolution of 10 nm has a size of approx. 32 GB with a single color. Larger screens reach PB size very quickly. Iterative variations based de-noise and de-blurr algorithms are needed to prevent artifacts at various stages of the image analysis [23]. Current systems are limited at both, processing and the intake of data. Data rates from a single nanoscope, in 3D, are in the range of several GB/s. A single threaded application is used for analysis.

The excellence cluster CellNetworks of the University of Heidelberg [24] addresses, among other, biomedical questions in the area of structural biology and structural cell biology. Contributing to an array of scale invariant imaging techniques research in cryo electron microscopy investigates 3D reconstruction of electron microscopic tomogram images using e.g. filtered backprojection and least squares solution of backprojection matrix [25]. The workflow produces almost 300 images with 18 GB data per tomogram. The setup is able to run 40 to 60 tomograms per day using 2 microscopes. 2D image analysis and 3D reconstruction of this data is only possible with a highly automated processing workflow and by moving part of the computation to special purpose nodes using GPUs.

Second generation single plane illumination microscopes (SPIM) [26] are able to reach very high resolutions of 3D images in only 30 seconds recording time. This technology allows recording over a long period collecting information about the influence of certain toxic environments to the development of a living zebra fish embryo. An observation time of 10 hrs results in datasets of approximately 7 TB which have to be transferred to the data facility, stored and analyzed. Analysis contains complex image processing algorithms like registration, segmentation and automatic feature extraction. To avoid bottlenecks resulting in huge delays in data analysis, challenges in high performance data management and data intensive computing have to be solved.

*E. Data and analysis in future large scale physics. The FAIR project*

The Helmholtzzentrum für Schwerionenforschung GmbH (GSI) [27] operates a large and in many aspects unique accelerator facility. Centered on GSI, an international structure named FAIR (Facility for Antiprotons and Ion Research) [28] will evolve. FAIR will generate beams of a previously unparalleled intensity and quality. In the final design FAIR consists of eight ring colliders with up to 1,100 meters in circumference, two linear accelerators and about 3.5 kilometer beam control tubes. The existing GSI accelerators serve as an injector. About 3000 scientists and engineers from more than 40 countries are already involved in the planning and development of the facility and its experiments. FAIR will support a wide variety of science cases: extreme states of matter using heavy ions (CBM), nuclear structure- and astrophysics (NUSTAR), hadron physics with antiprotons (PANDA), atomic and plasma physics as well as biological and material sciences (APPA).

The high beam intensities at FAIR constitute various challenges: very efficient accelerators, remote handling in activated areas, novel methods of cooling for the detectors, progress in collaborative computing and synergies between the various fields of research especially concerning the interaction with industry. The first beam is expected to be in 2018. This constitutes a crucial milestone in computing.

For two of the research topics of FAIR (CBM and PANDA) characteristics of the computing and data analysis are similar to the LHC at CERN that relies on trigger systems, but need further extensions. Trigger systems were necessary due to real-world limitations in data transport and processing bandwidth. At FAIR a novel triggerless detector read-out will be implemented, without conventional first-level hardware triggers, relying exclusively on event filters. This approach is much more flexible and adaptable to yet unforeseeable needs because the full detector information is available even in the first decision stage. Complex filtering and flexibility are the key enabling factors for the experiments at FAIR. So at FAIR the classical separation between data acquisition, trigger, and off-line processing is merging into a single, hierarchical data processing system, handling an initial data stream exceeding 1 TB/sec. The first layer of the system constitutes the first level event selector (FLES). The FLES implements a combination of specialized processing elements such as GPUs, CELL or FPGAs in combination with standard HPC clusters comprising compute nodes coupled by an efficient high-speed network. After the FLES the data stream fans out into an archival system and into the next distributed processing layers, which can be off-site.

Currently large-scale research infrastructures like e.g. LHC at CERN rely on on-site tier-0 data centers which perform the first-level processing after the last trigger step. Subsequently a considerably reduced amount of data is analyzed off-site in downstream data centers. In contrast to that FAIR will make use of a novel modular data processing paradigm using multi-site load balancing data centers. Several sites in the surrounding of GSI will be connected with a high-speed Metropolitan Area Network via fibre link allowing the off-loading of processing between the sites. Multi-site user access and corresponding security aspects are currently being investigated within the EU project CRISP [29]. That combined tier-0/1 system will be integrated in an international Grid/Cloud infrastructure. The FAIRGrid distributed computing infrastructure is currently implemented as two separate entities: PandaGrid [30] and CBMGrid. FAIRGrid uses the AliEn middleware [31], as developed by the ALICE experiment. The grid monitoring and data supervision are done via MonALISA. The basis of the software for simulation, reconstruction, and data analysis for the large FAIR experiments is FairRoot [32]. In order to meet peak demands for computing, it may be necessary to offload some of the computing tasks to public or community clouds. Thus, it becomes important to deploy and operate an infrastructure to compute these tasks on virtual machines in a quick and scalable way.

The exact amount of computing, storage and archiving required for FAIR depends on many factors but is certainly beyond the capacity of a single computer center. The required resources are dominated by CBM and PANDA. Current estimates for the sum of all experiments are 200.000 cores and 30 PB storage space for the first year of data taking. The load balancing distributed infrastructure must be developed as well as the seamless integration with existing grid and cloud infrastructures. The re-use of components from existing running high energy physics experiments poses specific challenges for data management and distributed computing that have to be solved within a few years.

## III. RELATED ACTIVITIES AND INFRASTRUCTURES

### A. GridKa

In 2002, the former Research Centre Karlsruhe (now KIT) started GridKa [2], providing compute power and storage for High Energy Physics experiments. Many physics experiments, e.g. CDF, D0, Babar and Compass, used GridKa for production while the LHC experiment collaborations already started to simulate their experiments and detectors. During this time, GridKa was participating in the EU-DataGrid project and later the LCG-1 and LCG-2 Grid projects.

In the following years the number of nodes and storage capacity was increased yearly to accommodate requirements of the particle physics community during which the LHC experiments developed and tested their distributed computing models. Numerous "Data and Service Challenges" have been performed to test these models and demonstrate the capabilities of the various participating Grid centers. In 2008, the Worldwide LHC Computing Grid [3] achieved the required target data rates and was able to demonstrate its readiness to store, distribute and compute the data to be produced by the LHC.

GridKa is one of 11 Tier-1 centers of the Worldwide LHC Computing Grid and one of four centers providing services and resources for all four of the large LHC experiments. It currently provides 130000 HEPSPEC'06 [33], approx. 11 PB of disk space and 17 PB of tape space. The disk and tape space is provisioned by several Grid enabled dCache [34] and XROOTD [35] storage systems, accessible from the wide area networks, capable of delivering I/O rates of several Gigabytes/s. The gLite File Transfer Service [36], controlling wide area data traffic to and from other Grid centers as well as workload management systems, information systems and file catalogues are provided as high-available Grid services.

### B. HPC Simulation Labs

At Karlsruhe Institute of Technology (KIT) and the Forschungszentrum Jülich (FZJ) Simulation Laboratories (SimLabs) for different research communities have been established. For example, the SimLab Energy at KIT focuses on models of chemical processes and of flows e.g. of hydrogen combustion and neutron physics, while the SimLab NanoMikro at KIT targets the simulation of nanostructured materials e.g. with density function theory. The other SimLabs at KIT are the SimLab Elementary and Astro-Particles and Climate and Environment. At FZJ the SimLabs Biology, Molecular Systems, Plasma Physics and Climate Science exist – the later one in close collaboration with the resp. SimLab at KIT.

A SimLab [5] is a community-oriented research and support group acting as interface between the supercomputing facilities and the university groups, institutes and communities. Its mission is to provide high level support in utilizing HPC facilities, especially at FZJ or KIT, by porting model systems to supercomputers and subsequently optimizing the performance of these models by enhancing their parallel scalability, load balancing and data access. The focus is support for High Performance Computing (HPC), but a SimLab also performs its own research together with the scientific communities. This work is typically done in close cooperation with scientific institutes through common projects.

The relationship between SimLabs and Data Life Cycle Labs in LSDMA is that in both instruments IT experts work in close collaboration with domain scientists. In the case of SimLabs on the particular challenges of enabling scientific communities to make efficient use of supercomputers for their simulation codes – in particular considering multi/many-core processors and Peta/Exaflop/s system performance.

### C. LSDF

The Large Scale Data Facility (LSDF) [4] was started at the end of 2009 at KIT with the aim of addressing the data management and processing requirements of several forthcoming data intensive experiments [37]. In particular, the High Throughput Microscopy experiments used for biological studies of zebra fishes and the Tomography beamline of the synchrotron radiation source ANKA, were planning to collect data amounting to several Petabytes in the coming years. Providing not only the storage capacity, but also the data management and analysis services was identified as a critical aspect where a unified Campus-wide approach would profit from many synergies, and relieve the communities from the administering a IT Infrastructure. Soon after the project started many scientific groups on the campus expressed their interest and joined. Currently the LSDF is providing storage through the GridFTP, NFS, and CIFS protocols [37], as well as an iRODS Datagrid [38]. The experimental data benefits from data management functionality if ingested with the tools developed within the project. Additionally, the data analysis can be carried out by means of the project's Cloud or Hadoop services [39].

A close collaboration of LSDMA and LSDF is natural, as LSDMA builds to a large extend on services and infrastructure such as LSDF provides. Currently, the LSDF is hosting around 1 PB of data and provides Hadoop and Cloud resources to scientific experiments from 17 KIT institutes. The number of scientists using these services lies in the range 80-100.

### D. EUDAT

The European Commission funds the European Data Infrastructure (EUDAT) [40] aiming at establishing a cost-efficient and high-quality collaborative data infrastructure throughout Europe. It addresses the needs and requirements of various scientific disciplines both in terms of capacity and capability. 25 partners from 13 countries focus on the development and operation of services for federating existing resources, which are provided by the EUDAT partners. Both FZJ and KIT are partners in LSDMA and EUDAT. Further services address common, baseline data services for a variety of communities, long-term archival with bit-stream persistence and integrity for a long-term and distinct referencing of data, proper up/download services for users as well as a federated security infrastructure for authentication and authorization.

A close collaboration of LSDMA and EUDAT is foreseen in which national services and processes are aligned with those on European level in order to enable scientific communities and boost the pan-European data-driven scientific discovery.

## IV. MANAGEMENT AND ANALYSIS OF SCIENTIFIC DATA - THE LSDMA PROJECT

After its proposal was accepted in late 2011, the project LSDMA started on January 1st, 2012. Its initial phase ends on December 31s, 2016, but the project is projected to be integrated into the sustainable Program oriented funding framework of the Helmholtz Association. The project partners for the initial phase are four Helmholtz Association research centers, namely DESY, FZJ, GSI and KIT, as well as six universities, namely HTW Berlin, Technical University of Dresden, University of Frankfurt, University of Hamburg, University of Heidelberg, University of Ulm, and the German climate research center DKRZ.

Reaching the fundamental goal of sustainably improving data analysis chains and data life cycles also depends on availability of data management components and their development. Standardized and generic tools have to be provided and have to be promptly researched, developed and established in a joint R&D program, run by data specialists and driven by user communities.

These two activities are reflected by the LSDMA project structure: several data life cycle labs (DLCLs) are closely connected to five of the six Helmholtz Association research fields enhanced with a data services integration team (DSIT), cf. Fig 2.

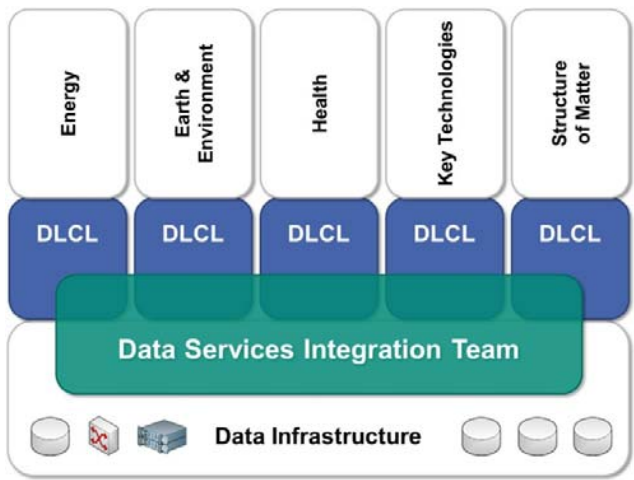

Figure 2. Principle idea of the LSDMA project

### A. Data Life Cycle Labs

At the core of the LSDMA initiative we find the community specific Data Life Cycle Labs. Each DLCL is associated with one or more initial topics:

- Energy: smart grids, battery research, fusion research
- Earth and Environment: climate models, environmental satellite data
- Health: virtual human brain map
- Key Technologies: synchrotron radiation, nanoscopy, high throughput microscopes, electron-microscope imaging techniques
- Structure of Matter: large instruments (ESFRI items Petra 3, X-Ray Free Electron Laser (XFEL), Facility for Antiprotons and Ions Research (FAIR), heavy ion research, elementary particle physics

In each Data Life Cycle Lab, a team of data scientists collaborates closely with scientist from the respective domain. They work together in joint R&D and iteratively optimize community-specific data life cycles, define and optimize data and meta-data formats, establish standardized data management techniques and improve ease of access and ease of use of the data locally, nationally and internationally. This cooperation will enhance the community collaboration by introducing new or selected existing data management standards, technologies and tools. A major and often underestimated effort of the R&D activity is carried out to support and to preserve data, methods and scientific insights in long-term, trustworthy and openly accessible archives.

The community-centric DLCL approach is to be rolled out for even more communities. As described in a concept of the Helmholtz Association [41], up to 13 additional DLCLs supporting communities that span the broad science spectrum of the Helmholtz Association from marine research to magnetic plasma are planned at Helmholtz centers all over Germany. The DLCLs are driven by the need of the communities, and are linked to topical and general data centers which in turn are connected in a federated research data infrastructure.

### B. Data Services Integration Team

All requirements and components of importance to more than one DLCL are taken over by DSIT. The team's R&D also includes the challenges identified in [42]. Such a cross-sectional approach is needed to efficiently handle large scale data management and data intensive analysis in the near future. For all of the science areas described above the cross sectional activity DSIT will specifically address the following topics:

- Distributed data management, storage and access
- Consolidation of different storage, access technologies and analysis infrastructure
- Metadata ontologies to identify data and its derivation over time
- Integration of user friendly graphic data processing workflow tools in the data life cycle
- Standardization of data and meta-data formats
- Data security
- High performance data analysis

DSIT is structured in four sub teams: Data-Intensive Computing and Applications (DIC), Migration, Preservation und Curation (MPC), Universal Data Access (UDA) and Storage System Design (SSD).

Further generic R&D activities were identified in the before mentioned Helmholtz concept by the scientific communities ranging from establishing worldwide data federations, improving data-intensive workflows, optimizing searching, indexing and meta-data methods up to high-performance data access, improved data aggregation and proper interfaces for platform integration.

All R&D in LSDMA is based on the communities' requirements for an optimized data life cycle. As this optimal data life cycle evolves with time, R&D in LSDMA has to adjust swiftly, dropping the development of specific software tools and replacing them with other tools.

*C. Time Line*

As the needs and requirements by the communities change over time, two major reports are fully updated each year. In spring, there is a detailed requirements analysis, composed by the DLCL on the bases of in-depth discussions with the community, which gives an extensive overview on the current and the aspired state of each data life cycle and summarizes the steps to be taken to achieve the latter. During summer, the project schedule for the upcoming 18 months is reviewed and adjusted to the needs described in the detailed requirement analysis and to the latest technical state of the art.

Starting the second year, there is a community forum every year, where communities exchange their experiences and articulate their requirements and wishes. Each spring, there is an all-hands meeting; in fall, there is an annual international LSDMA symposium, bringing together big data experts from all over the globe and LSDMA collaborators.

V. CONCLUSIONS

Recognizing data as source for new knowledge, the deployment of five Data Life Cycle Labs within the LSDMA project of the German Helmholtz Association is well underway. By combining the competence of several research centers, universities and communities in a novel superstructure, an accepted and efficient research and development platform for data management of the data life cycle is successfully established. The outlook towards a generic infrastructure in which all Helmholtz centers participate ensures a long term support for many scientific communities and gives the project a leading role in Germany.

The core building block is the Data Life Cycle Lab (DLCL) in which data experts closely work together on joint R&D topics with community scientists. Together they address the iterative optimization of community-specific data life cycles, the creation of services for simple access and use of data and local, national and international infrastructures for data storage, data processing and data archival as well as for international collaboration. The DLCLs are complemented with by a Data Services Integration Team (DSIT), where a specialized team of data experts work on generic, multi-community data management and analysis services and – together with the DLCLs – the integration of these services in the data life cycle of the scientific communities. The developed generic components and tools can be reused and applied for a growing range of scientific disciplines, allowing faster results and compliance with requirements for publishing results of scientific activities.

Already in its first year of existence, the project has gained considerable momentum in the scientific community and new communities are interested to establish DLCLs for their specific needs.

With its concept LSDMA facilitates scientific communities through advanced data exploration and preservation services with faster data extraction from original, derived and archived data, thereby immediately gaining scientific insight and enabling new societal knowledge in the future.

ACKNOWLEDGMENT

The authors wish to thank all people and institutions involved in defining and setting up the LSDMA project as well as the German Helmholtz Association for providing the funding.

REFERENCES

[1] J. Gray. eScience Talk at NRC-CSTB meeting Mountain View CA, 11 January 2007, http://research.microsoft.com/en-us/um/people/gray/talks/NRC-CSTB_eScience.ppt
[2] Grid Computing Centre Karlsruhe, http://www.gridka.de/
[3] Worldwide LHC Computing Grid, http://wlcg.web.cern.ch/
[4] A. García, S. Bourov, A. Hammad, J. van Wezel, B. Neumair, A. Streit, V. Hartmann, T. Jejkal, P. Neuberger, R. Stotzka. The Large Scale Data Facility: Data Intensive Computing for scientific experiments. Proceedings of The 12th IEEE International Workshop on Parallel and Distributed Scientific and Engineering Computing (PDSEC-11/IPDPS-11), IEEE Computer Society, 1467-1474 (2011), http://dx.doi.org/10.1109/IPDPS.2011.286
[5] N. Attig, R. Esser, P. Gibbon. Simulation Laboratories: An Innovative Community-Oriented Research and Support Structure. In: CGW'07, ACC CYFRON ET AGH, 2008, ISBN 978-83-915141-9-1, pages 1-9
[6] T. Hey, S. Tansley, K. Tolle. The Fourth Paradigm: Data-Intensive Scientific Discovery. Microsoft Research, ISBN 978-0982544204, http://research.microsoft.com/en-us/collaboration/fourthparadigm/
[7] M. Riese, F. Friedl-Vallon, R. Spang, P. Preusse, C. Schiller, L. Hoffmann, P. Konopka, H. Oelhaf, T. von Clarmann, M. Höpfner. GLObal limb Radiance Imager for the Atmosphere (GLORIA): Scientific objectives. Adv. Space Res., 36:989–995, 2005.
[8] M. Endemann. MIPAS instrument concept and performance. In Proceedings of ESAMS '99 – European Symposium on Atmospheric Measurements from Space, pages 29–43, 1999.
[9] ESA Envisat, http://www.esa.int/Envisat
[10] E. Guilyardi, V. Balaji, S. Callaghan, C. DeLuca, G. Devine, S. Denvil, R. Ford, C. Pascoe, M. Lautenschlager, B. Lawrence, L. Steenman-Clark, S. Valcke. The CMIP5 model and simulation documentation: a new standard for climate modelling metadata. CLIVAR Exchanges Special Issue No. 56, Vol. 16, No.2, May 2011
[11] F. Giorgi, C. Jones, G.R. Asrar. Addressing climate information needs at the regional level: The CORDEX framework. World Meteorological Organization Bulletin 58(3), 175–183, 2009
[12] M. Lautenschlager. Institutionalisierte „Data Curation Services", S. 149-156. Aus: Handbuch Forschungsdatenmanagement, Herausgegeben von Stephan Büttner, Hans-Christoph Hobohm, Lars Müller, BOCK + HERCHEN Verlag Bad Honnef 2011
[13] M. Greenwald, D. P. Schissel, J. R. Burruss, T. Fredian, J. Lister, J. Stillerman. Visions for Data Management and Remote Collaboration on ITER. Procceddings of the 10th International Conference on Accelerator and Large Experimental Physics Control Systems, Geneva, Switzerland (2005)
[14] International Thermonuclear Experimental Reactor, http://www.iter.org/
[15] Massoud Amin, S. and Wollenberg, B.F., Toward a smart grid: power delivery for the 21st century. Power and Energy Magazine 3/5, 34-41, IEEE
[16] M. Mültin, F. Allerding, H. Schmeck. Integration of electric vehicles in smart Homes - An ICT-based solution for V2G scenarios. IEEE PES Innovative Smart Grid Technologies, Washington DC USA Jan 2012.

[17] C. Daniel, J. O. Besenhard. Handbook of Battery Materials, Second Edition. 2011 Wiley-VCH Verlag GmbH & Co. KGaA

[18] http://www.fz-juelich.de/inm/inm-1/EN/Forschung/_docs/BrainMapping/BrainMapping_node.html

[19] K. Zilles, K. Amunts. Centenary of Brodmann's map – conception and fate. Nature Reviews Neuroscience 11(2): 139-145.

[20] The ANKA synchrotron facility, http://ankaweb.fzk.de/

[21] R. Stotzka, W. Mexner, T. dos Santos Rolo, H. Pasic, J. van Wezel, V. Hartmann, T. Jejkal, A. Garcia, D. Haas, A. Streit. Large Scale Data Facility for Data Intensive Synchrotron Beamlines. 13th International Conference on Accelerator and Large Experimental Physics Control Systems, 2011

[22] P. Lemmer, M. Gunkel, D. Baddeley, R. Kaufmann, A. Urich Y. Weiland, J. Reymann, P. Müller, M. Hausmann, C. Cremer. SPDM – light microscopy with single molecule resolution at the nanoscale. Appl. Phys. B 93: 1-12, 2008Müller, P.; Schmitt, E.; Jacob, A.; Hoheisel, J.; Kaufmann, R.; Cremer, C. & Hausmann, M. (2010), 'COMBO-FISH enables high precision localization microscopy as a prerequisite for nanostructure analysis of genome loci', Int J. Molec. Sci. 11, 4094-4105.

[23] N. Becherer, H. Jödicke, G. Schlosser, J. Hesser, F. Zeilfelder, R. Männer. On Soft Clipping of Zernike Moments for Deblurring and Enhancement of Optical Point Spread Functions. In 'Image Processing: Algorithms and Systems, Neural Networks, and Machine Learning, Proceedings of the SPIE', pp. 73-83.

[24] CellNetworks, http://www.cellnetworks.uni-hd.de/

[25] K. Schultheiss, J. Zach, B. Gamm, M. Dries, N. Frindt, R.R. Schröder, D. Gerthsen. New electrostatic phase plate for phase-contrast transmission electron microscopy and its application for wave-function reconstruction. Microsc Microanal, 16, pages 785-794

[26] A. Kobitski, J. C. Otte, B. Schäfer, U. Strähle, U. Nienhaus. Development of digital scanned laser light sheet microscopy with high spatial and temporal resolution. Accepted Presentation at the "Annual Meeting of the German Biophysical Society", 23-26 September 2012, Göttingen, Germany

[27] GSI Helmholtzzentrum für Schwerionenforschung GmbH, http://www.gsi.de/en/

[28] FAIR - Baseline Technical Report, Volume 1-6, September 2006, ISBN: 3-9811298-0-6, for a pdf version see: http://www.fair-center.eu/fair-users/publications/fair-publications.html

[29] CRISP http://www.crisp-fp7.eu/

[30] D. Protopopescu, K. Schwarz, PandaGrid - A Tool for Physics (PS36-2-471) Journal of Physics: Conference Series {\bf 331} (2011) 072028)

[31] ALiEn, http://alien2.cern.ch/

[32] FairRoot, http://fairroot.gsi.de/

[33] HEP-SPEC'06 Benchmark, http://w3.hepix.org/benchmarks/doku.php

[34] dCache, http://www.dcache.org/

[35] XRootD, http://xrootd.slac.stanford.edu/

[36] gLite File Transfer Services (FTS), https://www.gridpp.ac.uk/wiki/GLite_File_Transfer_Service

[37] M. Sutter, V. Hartmann, M. Götter, J. van Wezel, A. Trunov, T. Jejkal, R. Stotzka. File Systems and Access Technologies for the Large Scale Data Facility. In: Remote Instrumentation for eScience and Related Aspects, Springer, 2012, ISBN 978-1-4614-0508-5, pages 239-256, http://dx.doi.org/10.1007/978-1-4614-0508-5_16

[38] M. Hedges, T. Blanke, A. Hasan. Rule-based curation and preservation of data: A data grid approach using iRODS. Future Generation Computer Systems, Vol. 25 Iss. 4, Elsevier, 2008, pages 446-452, http://dx.doi.org/10.1016/j.future.2008.10.003

[39] T. White. Hadoop: The Definitive Guide. O'Reilly, 2010, ISBN 978-1449389734

[40] EUDAT, http://www.eudat.eu/

[41] Concept for improving Supercomputing and Big Data in the German Helmholtz Association, in German, not yet publicly available

[42] e-IRG-HLG (High Level Expert Group on Scientific Data). Riding the Wave: How Europe can gain from the rising tide of scientific data. European Commission, 2010, http://cordis.europa.eu/fp7/ict/e-infrastructure/docs/hlg-sdi-report.pdf